\documentclass[runningheads]{llncs}
\usepackage[T1]{fontenc}
\usepackage{graphicx}
\usepackage{color}
\usepackage{subfig}
\usepackage{amsmath}
\usepackage{booktabs}
\usepackage{hyperref}
\usepackage[capitalize,noabbrev]{cleveref}
\usepackage[draft]{todonotes}
\usepackage[inkscapearea=page]{svg}

\begin{document}
\title{DQNC2S: DQN-based Cross-stream Crisis event Summarizer}
\titlerunning{DQNC2S: DQN-based Cross-Stream Crisis event Summarizer}
%

\author{Daniele Rege Cambrin\orcidID{0000-0002-5067-2118} \and
Luca Cagliero\orcidID{0000-0002-7185-5247} \and
Paolo Garza\orcidID{0000-0002-1263-7522}}

\authorrunning{D. Rege Cambrin et al.}

\institute{Politecnico di Torino, Torino, Italy\\
\email{\{daniele.regecambrin,luca.cagliero,paolo.garza\}@polito.it}}

%
%
%
\maketitle              
\begin{abstract}
Summarizing multiple disaster-relevant data streams simultaneously is particularly challenging as existing Retrieve\&Re-ranking strategies suffer from the inherent redundancy of multi-stream data and limited scalability in a multi-query setting. 
This work proposes an online approach to crisis timeline generation based on weak annotation with Deep Q-Networks. It selects on-the-fly the relevant pieces of text 
without requiring neither human annotations nor content re-ranking. This makes the inference time independent of the number of input queries.
The proposed approach also incorporates a redundancy filter into the reward function to effectively handle cross-stream content overlaps.
The achieved ROUGE and BERTScore results are superior to those of best-performing models on the CrisisFACTS 2022 benchmark. 


\end{abstract}

\section{Introduction}
Automating the extraction of valuable information from large streams of social and news data is particularly relevant to crisis 
management as could become an active support to disaster-response personnel~\cite{social_emergency,social_media_crisis}.
The problem of generating crisis event summaries of time-evolving data streams has already been addressed by the
Temporal Summarization~\cite{temp-summ} (TS) and Incident Stream~\cite{trec-is} (IS) challenges, 
where the goals are to extract variable-length summaries for a particular event on a set day (TS) 
and to classify and prioritize information in a single social stream (IS), respectively. 
More recently, the CrisisFACTS task~\cite{crisisfacts} has focused on retrieving daily timelines from social/news data streams.
The main challenges are (1) The contemporary presence of multiple streams of
crisis-relevant data (Twitter, Reddit, Facebook, and News), which makes the input collection particularly redundant and complex to summarize; 
(2) The lack of human annotations on text relevance and topic-level clusters, which limits the scope of supervised techniques;
(3) A list of queries about information needed by emergency responders for each emergency event, which increases the complexity compared to single-query tasks. 
State-of-the-art approaches to the CrisisFACTS task rely on ColBERT\cite{colbert} (ohmkiz~\cite{ohmkiz}) or on the established Retrieve\&Re-ranking approach (unicamp~\cite{unicamp} and ohmkiz~\cite{ohmkiz}). This last one entails retrieving the pieces of text relevant to an input query first and then re-rank them.
The retrieval step is commonly driven by text relevance scoring functions stored in index structures (e.g.,BM25~\cite{bm25}, Dense Passage Retriever~\cite{dense-passage-retrieval}) whereas the re-ranking step is commonly based on neural models such as ColBERT~\cite{colbert}, DRMM~\cite{drmm}, and Conv-KNRM~\cite{conv_knrm}. Both end-to-end ColBERT and Retrieve\&Rerank approaches scale linearly with the number of input queries and are not designed for online text processing. Furthermore, it is necessary to incorporate a filtering stage after the re-ranking phase thus retrieving large volumes of redundant content in a multi-stream scenario.  

This work proposes a new DQN-based Cross-Stream Crisis event Summarizer. It overcomes the main issues of existing approaches to CrisisFACTS by adopting Deep Reinforcement Learning (DRL). DRL has proved to be effective in both recommender systems~\cite{drl_reccomender} and text summarization~\cite{abstractive-rl}. 
The key contributions of the present work are enumerated below.

\begin{itemize}
\item {\bf RL approach for online text retrieval.} Unlike state-of-the-art approaches (e.g.,~\cite{unicamp}), 
we rely on DRL. Hence, we do not need to index crisis-relevant data for efficient content retrieval. Conversely, we retrieve relevant texts on-the-fly from the input streams without any human supervision. 
\item {\bf Early redundancy filtering.} The RL reward function already incorporates a redundancy filter to reduce the amount of data processed after retrieval. This improves the efficiency and reduces the complexity of the timeline generation method.
\item {\bf Efficient multi-query setting.} Unlike existing approaches to CrisisFACTS, our approach inherently supports multiple queries simultaneously. This allows the efficient generation of timelines of multi-faceted crisis events, where each query refers to a complementary facet.
\end{itemize}

\noindent Our approach performs better than the state of the art on the CrisisFACTS benchmark. The project code is publicly available for research purposes\footnote{\url{https://github.com/DarthReca/crisis-dqn} Latest access: October 2023}.

\section{Problem Statement}
\label{sec:problem}
We address the CrisisFACTS 2022 task first proposed in~\cite{crisisfacts}. The goal is to summarize multiple streams of social and news data describing the same short-lasting crisis event. 
Let $E$ be a set of crisis events about different hazards (see \cref{tab:events}).
Each event lasts at least two days within the reference time period $T$ and is described by a set $S$ of streams of textual content, i.e., tweets, Reddit messages, news, and Facebook posts\footnote{Due to the restrictions in force to the Facebook data crawler, similar to most peers hereafter we will disregard this stream.}. 
Stream data consists of timestamped pieces of text $te^t_s$, where $s \in S$ and $t \in T$. 
The task focuses on a set $Q_e$ of event queries for each $e \in E$ that are of interest to emergency responders. 
Given $e$, $S$, and $Q_e$, the CrisisFACTS task aims at generating a summary of $S$ that consists of daily crisis timelines reporting the top stories related to $Q_e$ on $e$. On each timestamp (day) $t$, the timeline consists of a shortlist of \textit{facts}, sorted by decreasing importance, and described by one or more pieces of text $te^t_s$.



\begin{table}[t!]
    \centering
    \footnotesize
    \caption{CrisisFACTS 2022 events}
    \label{tab:events}
\begin{tabular}{@{}l|llll@{}}
\toprule
Event ID & Name                      & Queries & Texts  & Days \\ \midrule
001      & Lilac Wildfire 2017       & 52      & 45578  & 9    \\
002      & Cranston Wildfire 2018    & 52      & 25172  & 6    \\
003      & Holy Wildfire 2018        & 52      & 25482  & 6    \\
004      & Hurricane Florence 2018   & 51      & 180286 & 15   \\
005      & Maryland Flood 2018       & 48      & 37598  & 4    \\
006      & Saddleridge Wildfire 2019 & 52      & 34480  & 4    \\
007      & Hurricane Laura 2020      & 51      & 52561  & 2    \\
008      & Hurricane Sally 2020      & 51      & 67632  & 8    \\ \bottomrule
\end{tabular}
\end{table}

\section{Methodology}
\label{sec:methodology}
We present the DQN-based Cross-Stream Crisis event Summarizer (DQNC2S, in short).
It encompasses a three-step process. First, we annotated text candidates based on 
established extractive QA models to guide the training of the DQN. 
Secondly, our approach relies on Deep Reinforcement Learning to online retrieve relevant yet non-redundant content
suitable for daily summaries. Finally, it applies topic modeling and abstractive summarization to synthesize and rephrase the retrieved texts.


\paragraph{Weak Annotation}
The use of extractive Question Answering (QA) models has proved effective in supporting the generation  
of crisis-related event descriptions~\cite{ohmkiz}. Inspired by the recent use of SQUAD~\cite{SQuAD2018} for evidence estimation, 
we perform a weak annotation step to create an importance score for each query-text pair.
Specifically, given a pair of query $q_i \in Q_e$ and a set $S$ of multiple streams related to crisis event $e$, 
we leverage Electra\footnote{\url{https://huggingface.co/deepset/electra-base-squad2} Latest access: October 2023} and LongFormer\footnote{\url{https://huggingface.co/allenai/longformer-large-4096-finetuned-triviaqa}  Latest access: October 2023} to generate tuples $<q_i,te^t_s,CF>$, where $CF$ is the mean confidence score associated by the QA models to the query-text pair. 
To avoid local model inaccuracies, a tuple is generated only when both QA models provide an answer with a minimal confidence level of 80\%. 
Each text gets a score $Sc$ equal to the number of queries it provides an answer to.

\paragraph{DQN-based Text Retrieval}
Our retriever is based on a Deep Q-Network~\cite{dqn}, which interacts with an environment designed to work online and driven by a single parameter, i.e., the maximum number of retrievable texts.
The action space is binary, i.e., $A = 0$ when the text is kept or $A = 1$ when it is discarded.
The observation space of size 770 comprises the BERT embedding of the current text (size 768), 
the remaining percentage of texts that can be chosen (size 1), and the maximum similarity between the current text and the other kept texts (size 1). 
The reward function $\mathcal{R}$ is defined in \cref{eq:reward}.
Specifically, if $Sc = 0$, the text should be discarded as unlikely to be informative.
If kept, the reward is -5 (value chosen after reward shaping); otherwise, the value is set to 1.
If the text has a non-zero score, it should be kept only if it is dissimilar enough to any previously selected texts (according to the cosine similarity between the corresponding embeddings).
We leverage a normalized score $N_{Sc}$=$\frac{Sc}{|Q_e|}$ to deal with a variable number of queries per event and then re-scale the product by the maximum similarity score $Si_m$.

\begin{equation}[htb]
    \mathcal{R} = 
    \begin{cases}
        -5 & \text{if } Sc = 0 \land A = 0 \\
        1  & \text{if } Sc = 0 \land A = 1 \\
        N_{Sc} - N_{Sc} * Si_m & \text{if } Sc > 0 \land A = 0 \\
        - (N_{Sc} - N_{Sc} * Si_m) & \text{if } Sc > 0 \land A = 1
    \end{cases}
    \label{eq:reward}
\end{equation}





\paragraph{Topic Modeling and Abstraction}
The outcomes of extractive summarization strategies are often suboptimal 
when the summary consists of several fragments of 
(potentially incoherent) text, especially while dealing with social data like tweets or Reddit posts. 
To cope with this issue, we explore the use of state-of-the-art topic modeling and abstractive summarization techniques, i.e., BERTopic~\cite{bertopic} and BART-CNN~\cite{bartcnn}.
The former technique groups the retrieved content into subtopics and provides a more faceted description of each fact.
The latter reformulates the original text into a concise and more readable form. 
DQNC2S-T, DQNC2S-A, and DQNC2S-T+A are the system variants incorporating either topic modeling, or abstraction, or a combination of the above. 

To generate the crisis event timeline, we produce daily facts. Each fact $F$ corresponds to a different topic and is described by one or more pieces of text. 
The importance score of text $te_s^t$ of fact $F$ is $I_{te_s^t,F} = Q_T - Q_D$, where $Q_T$ is the Q-value for taking the text, whereas $Q_D$ is the Q-value for discarding it. 
The higher the gap between the taking/discarding expected rewards, the higher the relevance of the text to the fact description. 

\section{Experiments}
\label{sec:experiments}

\paragraph{Dataset}
The CrisisFACTS dataset consists of multi-stream data annotated with
ground truth summaries extracted from Wikipedia, ICS-209 All-Hazards Dataset~\cite{ics209plus,ics}, and NIST annotation.

\paragraph{Evaluation metrics} In compliance with~\cite{crisisfacts}, we evaluated the results in terms of Rouge-2 F1-Score and BERT-Score metrics. They respectively quantify summary coherence from syntactic and semantic perspectives~\cite{AntonyASKDRPB23}. To make a fair summary comparison, each system shortlists a list of top-$k$ facts in order of decreasing importance, where $k$ is specified by the NIST assessors. Since our work also addresses efficiency-related aspects, we consider the execution time (expressed in seconds), too.

\paragraph{Baselines}
We shortlist the best-performing methods according to the results of the CrisisFACTS 2022 challenge, i.e., unicamp~\cite{unicamp} and ohmkiz \cite{ohmkiz}. They are both Retrieve\&Re-ranking strategies.  
We also consider a state-of-the-art end-to-end model tested by \cite{ohmkiz}, i.e., the ColBERT-based approach~\cite{colbert}.
We also include the baselines provided by the CrisisFACTS organizers~\cite{crisisfacts} for completeness.

\begin{figure}[tb]
    \centering
    \subfloat[Mean percentage of taken\label{fig:taken}]{\includegraphics[width=0.42\linewidth]{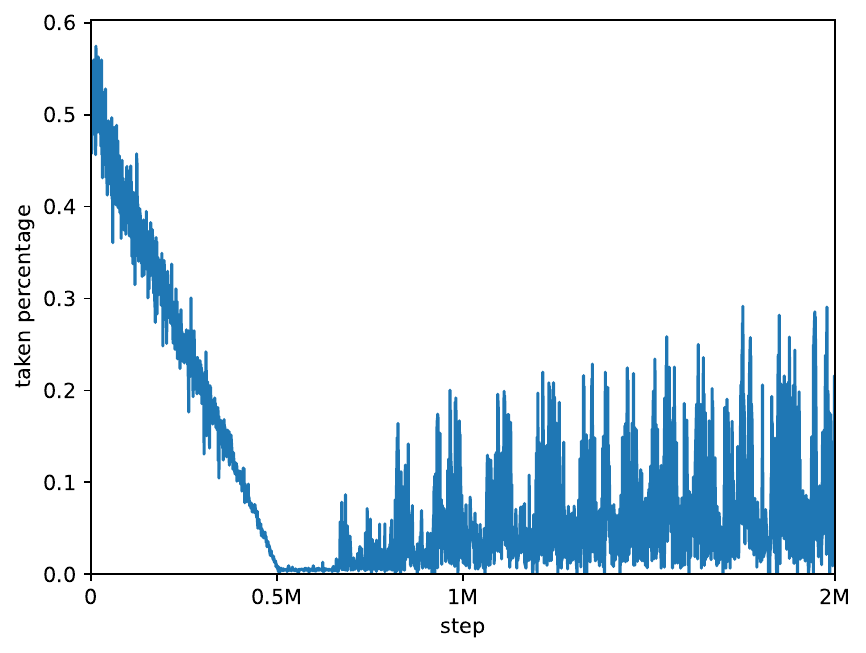}}
    \hfill
    \subfloat[Mean number of selected text\label{fig:selected}]{\includegraphics[width=0.42\linewidth]{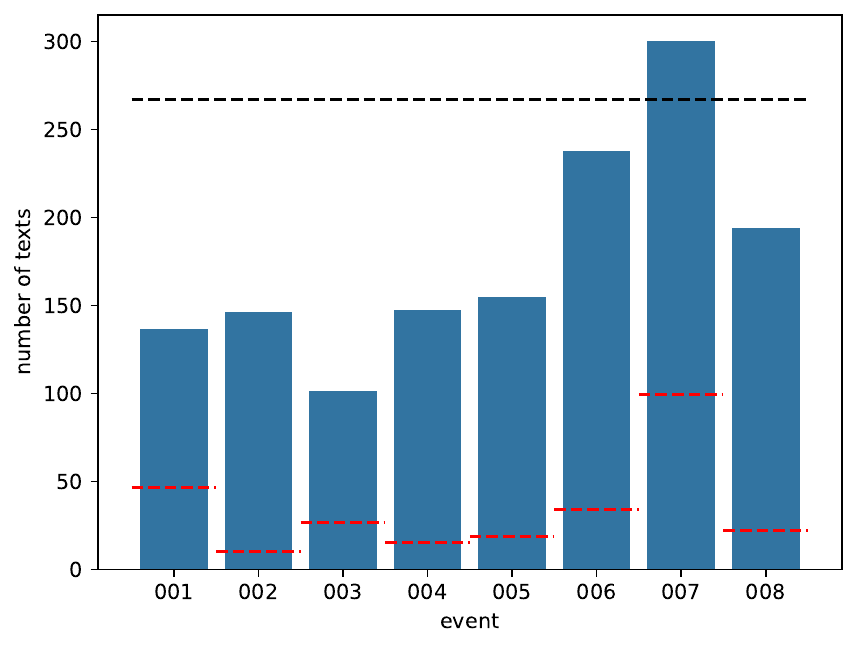}}
    \caption{Mean percentage of "take" action per episode during training (a) and the mean number of retrieved text per event (b). The red lines indicate the mean value and the black line the maximum number of daily facts according to the NIST annotation.}
    \label{fig:training_metrics}
\end{figure}

\paragraph{Setup}
We run the experiments on an Intel(R) Core(TM) i9-10980XE CPU and P5000 and A6000 GPUs.
The Q-Network comprises a mpnet-base-v2 version of SentenceBERT~\cite{sentence-bert} and three linear layers. 
We employ an Adam optimizer with a constant learning rate of 1e-3 and weight decay of 1e-4. To test the retrieve over multiple events, we perform an 8-fold cross-validation. In compliance with the empirical distribution of the number of facts per day (see~\cref{fig:training_metrics}), we set the maximum number of handled texts to 300.  

\begin{table*}[tb]
    \centering
    \caption{Comparison of mean Rouge-2 (R2) and BERT-Score (BS) F1-scores.}
    \label{tab:result_comparison}
    \resizebox{\linewidth}{!}{
\begin{tabular}{@{}l|cc|cc|cc||cc@{}}
\cmidrule(l){2-9}
                        & \multicolumn{2}{c|}{\textbf{ICS}}          & \multicolumn{2}{c|}{\textbf{NIST}}         & \multicolumn{2}{c||}{\textbf{Wikipedia}}    &  \multicolumn{2}{c}{\textbf{Mean}}     
                        \\ \midrule
 \textbf{Method }                 & R2              & BS              & R2              & BS              & R2              & BS              & R2              & 
 BS 
 \\ \midrule
baseline.run1           & 0.0418          & 0.4432          & 0.1326          & 0.5565          & 0.0281          & 0.5296          & 0.0675          & 0.5098          \\
baseline.run2           & 0.0428          & 0.4427          & 0.1308          & 0.5565          & 0.0281          & 0.5274          & 0.0672          & 0.5089          \\
ohm\_kiz.ColBERT        & 0.0497          & 0.4500          & 0.1386          & 0.5460          & 0.0307          & 0.5423          & 0.0730          & 0.5128          \\
ohm\_kiz.QACrisis       & 0.0464          & 0.4432          & 0.1471          & 0.5642          & 0.0337          & 0.5448          & 0.0757          & 0.5174          \\
ohm\_kiz.QAasnq         & 0.0507          & 0.4477          & 0.1468          & 0.5628          & 0.0362          & 0.5646          & 0.0779          & 0.5250           \\
unicamp.NM2             & \textbf{0.0581} & 0.4591          & 0.1338          & 0.5573          & 0.0281          & 0.5321          & 0.0733          & 0.5162          \\
unicamp.NM1             & \textbf{0.0581} & 0.4591          & 0.1338          & 0.5573          & 0.0281          & 0.5321          & 0.0733          & 0.5162          \\ \midrule
\text{DQN2CS}         & 0.0406          & 0.4554          & \textbf{0.1540} & \textbf{0.5715} & 0.0402          & 0.5516          & 0.0783          & 0.5262          \\
\text{DQN2CS-T}   & 0.0513          & 0.4579          & 0.1450          & 0.5667          & 0.0317          & 0.5538          & 0.0760          & 0.5261          \\
\text{DQN2CS-A}   & 0.0412          & \textbf{0.4596} & 0.1538          & 0.5706          & 0.0394          & 0.5538          & 0.0781          & 0.5280           \\
\text{DQN2CS-T+A} & 0.0452          & 0.4560          & 0.1515          & 0.5709          & \textbf{0.0420} & \textbf{0.5707} & \textbf{0.0796} & \textbf{0.5325} \\ \bottomrule
\end{tabular}
}
\end{table*}

\paragraph{Quantitative results overview}
As shown in \cref{tab:result_comparison}, DQNC2S performs best in terms of BERTScore for all the summary types, showing maximal semantic similarity with the reference crisis timelines. Its performance is best regarding ROUGE scores for all summaries except for ICS. According to the task organizers~\cite{crisisfacts}, the actual incident summaries in ICS are written for the public and from a historical perspective, not for the utility of emergency-response personnel. Thus, the syntactic n-gram overlap is likely less explanatory. 
DQNC2S-T+A turns out to be, on average, the most effective one. It is particularly effective in generating Wikipedia-like summaries, which provide topic-specific event insights.
According to the t-test of statistical significance ($p < 0.1$), our overall performance on CrisisFACTS is superior to the state of the art.

\paragraph{Inference time.}
DQNC2S decides if a piece of text is relevant in $0.0296 \pm 0.0037$ seconds for all the $N$ queries. 
Instead, unicamp~\cite{unicamp} requires approximately $N \cdot 0.0752 \pm 0.0380$ seconds just for the reranking step (disregarding the 
index creation and the usage of GPT-3).
Similarly, ohmkiz \cite{ohmkiz} requires around $N \cdot 0.0293 \pm 0.0190$ for text re-ranking.
The results confirm the higher efficiency of our strategy compared to state-of-the-art methods. 

\paragraph{DQN training and testing.}
\cref{fig:taken} shows the mean percentage of "take" action per episode. The system exploration phase proceeds until the 500000-th step, in which the DQN learns how to properly shortlist the relevant pieces of text. In the exploitation phase, the system learns to take all possible useful texts to maximize the reward, stabilizing its trend. Looking at the number of texts shortlisted at inference time in \cref{fig:selected}, the system rarely fills the whole pool. The correlation between the mean number of expected (red lines) and retrieved (blue bars) daily texts is relatively high (around $0.74$).

\section{Conclusion and Future Works}
\label{sec:conclusion}
In this work, we proposed a DRL approach to generate crisis timelines from cross-stream data. We tackled the limitations of existing 
strategies in
(1) Online processing of cross-stream data, avoiding ad-hoc indexing strategies for content retrieval;
(2) Extracting salient crisis-relevant information by early filtering redundant content during the text retrieval stage;    
(3) Efficiently handling multiple event queries, with an inference time independent of the number of input queries. 
We achieved +0.0063 R2 and +0.0151 BERTScore average improvements compared to the best-performing existing solutions. 
In future works, we plan to leverage Large Language Models to enrich the contextual representation of the crisis events. 

\section{Copyright statement}
This preprint has not undergone peer review (when applicable) or any post-submission improvements or corrections

%
%
%
\bibliographystyle{splncs04}
\bibliography{bibliography}

\begin{thebibliography}{10}
\providecommand{\url}[1]{\texttt{#1}}
\providecommand{\urlprefix}{URL }
\providecommand{\doi}[1]{https://doi.org/#1}

\bibitem{drl_reccomender}
Afsar, M.M., Crump, T., Far, B.: Reinforcement learning based recommender systems: A survey. ACM Comput. Surv.  \textbf{55}(7) (dec 2022). \doi{10.1145/3543846}, \url{https://doi.org/10.1145/3543846}

\bibitem{AntonyASKDRPB23}
Antony, D., Abhishek, S., Singh, S., Kodagali, S., Darapaneni, N., Rao, M., Paduri, A.R., BG, S.: A survey of advanced methods for efficient text summarization. In: 13th {IEEE} Annual Computing and Communication Workshop and Conference, {CCWC} 2023, Las Vegas, NV, USA, March 8-11, 2023. pp. 962--968. {IEEE} (2023). \doi{10.1109/CCWC57344.2023.10099322}, \url{https://doi.org/10.1109/CCWC57344.2023.10099322}

\bibitem{temp-summ}
Aslam, J.A., Diaz, F., Ekstrand-Abueg, M., McCreadie, R., Pavlu, V., Sakai, T.: Trec 2014 temporal summarization track overview. In: TREC (2015)

\bibitem{abstractive-rl}
Chen, Y.C., Bansal, M.: Fast abstractive summarization with reinforce-selected sentence rewriting. In: Proceedings of the 56th Annual Meeting of the Association for Computational Linguistics (Volume 1: Long Papers). pp. 675--686. Association for Computational Linguistics, Melbourne, Australia (Jul 2018). \doi{10.18653/v1/P18-1063}, \url{https://aclanthology.org/P18-1063}

\bibitem{conv_knrm}
Dai, Z., Xiong, C., Callan, J., Liu, Z.: Convolutional neural networks for soft-matching n-grams in ad-hoc search. In: Proceedings of the Eleventh ACM International Conference on Web Search and Data Mining. p. 126–134. WSDM '18, Association for Computing Machinery, New York, NY, USA (2018). \doi{10.1145/3159652.3159659}, \url{https://doi.org/10.1145/3159652.3159659}

\bibitem{ics209plus}
Denis, L.S., Mietkiewicz, N., Short, K., Buckland, M., Balch, J.: {ICS-209-PLUS - An all-hazards dataset mined from the US National Incident Management System 1999-2014}  (1 2020). \doi{10.6084/m9.figshare.8048252.v14}, \url{https://figshare.com/articles/dataset/ICS209-PLUS_Cleaned_databases/8048252}

\bibitem{bertopic}
Grootendorst, M.: Bertopic: Neural topic modeling with a class-based tf-idf procedure. arXiv preprint arXiv:2203.05794  (2022)

\bibitem{drmm}
Guo, J., Fan, Y., Ai, Q., Croft, W.B.: A deep relevance matching model for ad-hoc retrieval. In: Proceedings of the 25th ACM International on Conference on Information and Knowledge Management. p. 55–64. CIKM '16, Association for Computing Machinery, New York, NY, USA (2016). \doi{10.1145/2983323.2983769}, \url{https://doi.org/10.1145/2983323.2983769}

\bibitem{dense-passage-retrieval}
Karpukhin, V., Oguz, B., Min, S., Lewis, P., Wu, L., Edunov, S., Chen, D., Yih, W.t.: Dense passage retrieval for open-domain question answering. In: Proceedings of the 2020 Conference on Empirical Methods in Natural Language Processing (EMNLP). pp. 6769--6781. Association for Computational Linguistics, Online (Nov 2020). \doi{10.18653/v1/2020.emnlp-main.550}, \url{https://aclanthology.org/2020.emnlp-main.550}

\bibitem{colbert}
Khattab, O., Zaharia, M.: {ColBERT}: Efficient and effective passage search via contextualized late interaction over bert. p. 39–48. SIGIR '20, Association for Computing Machinery, New York, NY, USA (2020). \doi{10.1145/3397271.3401075}, \url{https://doi.org/10.1145/3397271.3401075}

\bibitem{bartcnn}
Lewis, M., Liu, Y., Goyal, N., Ghazvininejad, M., Mohamed, A., Levy, O., Stoyanov, V., Zettlemoyer, L.: {BART:} denoising sequence-to-sequence pre-training for natural language generation, translation, and comprehension. CoRR  \textbf{abs/1910.13461} (2019), \url{http://arxiv.org/abs/1910.13461}

\bibitem{social_emergency}
Lorini, V., Castillo, C., Peterson, S., Rufolo, P., Purohit, H., Pajarito, D., de~Albuquerque, J.P., Buntain, C.: Social media for emergency management: Opportunities and challenges at the intersection of research and practice. In: 18th international conference on information systems for crisis response and management. pp. 772--777 (2021)

\bibitem{crisisfacts}
McCreadie, R., Buntain, C.: Crisisfacts: Buidling and evaluating crisis timelines pp. 320--339 (2023). \doi{http://dx.doi.org/10.59297/JVQZ9405}

\bibitem{trec-is}
McCreadie, R., Buntain, C., Soboroff, I.: Trec incident streams: Finding actionable information on social media  (2019)

\bibitem{dqn}
Mnih, V., Kavukcuoglu, K., Silver, D., Rusu, A.A., Veness, J., Bellemare, M.G., Graves, A., Riedmiller, M., Fidjeland, A.K., Ostrovski, G., et~al.: Human-level control through deep reinforcement learning. nature  \textbf{518}(7540),  529--533 (2015)

\bibitem{unicamp}
Pereira, J., Fidalgo, R., Lotufo, R., Nogueira, R.: Using neural reranking and gpt-3 for social media disaster content summarization. In: Proceedings of the 31st Text Retrieval Conference (TREC 2022) Ian Soboroff and Angela Ellis, eds. (2023), \url{https://trec.nist.gov/pubs/trec31/papers/NM.unicamp.R.pdf}

\bibitem{SQuAD2018}
Rajpurkar, P., Jia, R., Liang, P.: Know what you don't know: Unanswerable questions for squad. In: Gurevych, I., Miyao, Y. (eds.) Proceedings of the 56th Annual Meeting of the Association for Computational Linguistics, {ACL} 2018, Melbourne, Australia, July 15-20, 2018, Volume 2: Short Papers. pp. 784--789. Association for Computational Linguistics (2018). \doi{10.18653/v1/P18-2124}, \url{https://aclanthology.org/P18-2124/}

\bibitem{sentence-bert}
Reimers, N., Gurevych, I.: Making monolingual sentence embeddings multilingual using knowledge distillation. In: Proceedings of the 2020 Conference on Empirical Methods in Natural Language Processing. Association for Computational Linguistics (11 2020), \url{https://arxiv.org/abs/2004.09813}

\bibitem{bm25}
Robertson, S., Zaragoza, H.: The probabilistic relevance framework: Bm25 and beyond. Found. Trends Inf. Retr.  \textbf{3}(4),  333–389 (apr 2009). \doi{10.1561/1500000019}, \url{https://doi.org/10.1561/1500000019}

\bibitem{social_media_crisis}
Saroj, A., Pal, S.: Use of social media in crisis management: A survey. International journal of disaster risk reduction  \textbf{48},  101584 (2020), \url{https://api.semanticscholar.org/CorpusID:218780990}

\bibitem{ohmkiz}
Seeberger, P., Riedhammer, K.: Combining deep neural reranking and unsupervised extraction for multi-query focused summarization. arXiv preprint arXiv:2302.01148  (2023)

\bibitem{ics}
St.~Denis, L.A., Short, K.C., McConnell, K., Cook, M.C., Mietkiewicz, N.P., Buckland, M., Balch, J.K.: all-hazards dataset mined from the us national incident management system 1999--2020. Scientific data  \textbf{10}(1), ~112 (2023)

\end{thebibliography}

\end{document}